\documentclass[]{spie}  %>>> use for US letter paper
\usepackage{amsmath,amsfonts,amssymb}
\usepackage{graphicx}
\usepackage[colorlinks=true, allcolors=blue]{hyperref}
\usepackage[]{cite}

\title{MagAO-X Phase II Upgrades: Implementation and First On-Sky Results of a New Post-AO 1000 Actuator Deformable Mirror}

\author[a,b,j]{Jay K. Kueny}
\author[a]{Kyle Van Gorkom}
\author[a,b,j]{Maggie Kautz}
\author[a,h]{Sebastiaan Haffert}
\author[a]{Jared R. Males}
\author[d]{Alex Hedglen}
\author[a]{Laird Close}
\author[a,b,j]{Eden McEwen}
\author[a,j]{Jialin Li}
\author[e]{Joseph D. Long}
\author[a]{Warren Foster}
\author[a]{Logan Pearce}
\author[f]{Avalon McLeod}
\author[g]{Jhen Lumbres}
\author[i]{Lauren Schatz}
\author[h]{Olivier Guyon}
\author[a,b]{Joshua Liberman}
\affil[a]{Steward Observatory, University of Arizona, Tucson, 933 N Cherry Ave, Tucson, AZ 85721, USA}
\affil[b]{James C. Wyant College of Optical Sciences, University of Arizona, 1630 E. University Blvd., Tucson, AZ 85721, USA}
\affil[c]{Leiden Observatory, Leiden University, PO Box 9513, 2300 RA Leiden, The Netherlands}
\affil[d]{Northrop Grumman in Rolling Meadows, IL}
\affil[e]{Center for Computational Astrophysics, Flatiron Institute, NY}
\affil[f]{Draper Laboratory in Cambridge, MA}
\affil[g]{Northrop Grumman in Pasadena, CA}
\affil[h]{Subaru Telescope, National Observatory of Japan, Hilo, HI}
\affil[i]{Starfire Optical Range, Kirtland Air Force Base in Albuquerque, NM}
\affil[j]{National Science Foundation Graduate Research Fellow}

\authorinfo{Further author information: (Send correspondence to J.K.K)\\J.K.K.: E-mail: jkueny@arizona.edu \\ }

\begin{document} 
\maketitle

\begin{abstract}
MagAO-X is the extreme coronagraphic adaptive optics (AO) instrument for the 6.5-meter Magellan Clay telescope and is currently undergoing a comprehensive batch of upgrades. One innovation that the instrument features is a deformable mirror (DM) dedicated for non-common path aberration correction (NCPC) within the coronagraph arm.  We recently upgraded the 97 actuator NCPC DM with a 1000 actuator Boston Micromachines \textit{Kilo-DM} which serves to (1) correct non-common path aberrations which hamper performance at small inner-working angles, (2) facilitate focal-plane wavefront control algorithms (e.g., electric field conjugation) and (3) enable 10 kHz correction speeds (up from 2 kHz) to assist post-AO, real-time low-order wavefront control. We present details on the characterization and installation of this new DM on MagAO-X as part of our efforts to improve deep contrast performance for imaging circumstellar objects in reflected light. Pre-installation procedures included use of a Twyman-Green interferometer to build an interaction matrix for commanding the DM surface, in closed-loop, to a flat state for seamless integration into the instrument. With this new NCPC DM now installed, we report on-sky results from the MagAO-X observing run in March---May 2024 for the Focus Diversity Phase Retrieval and implicit Electric Field Conjugation algorithms for quasistatic speckle removal and in-situ Strehl ratio optimization, respectively.
  
\end{abstract}


\section{INTRODUCTION}
\label{sec:intro}  % \label{} allows reference to this section

\begin{figure}
   \begin{center}
   \includegraphics[width=0.8\linewidth]{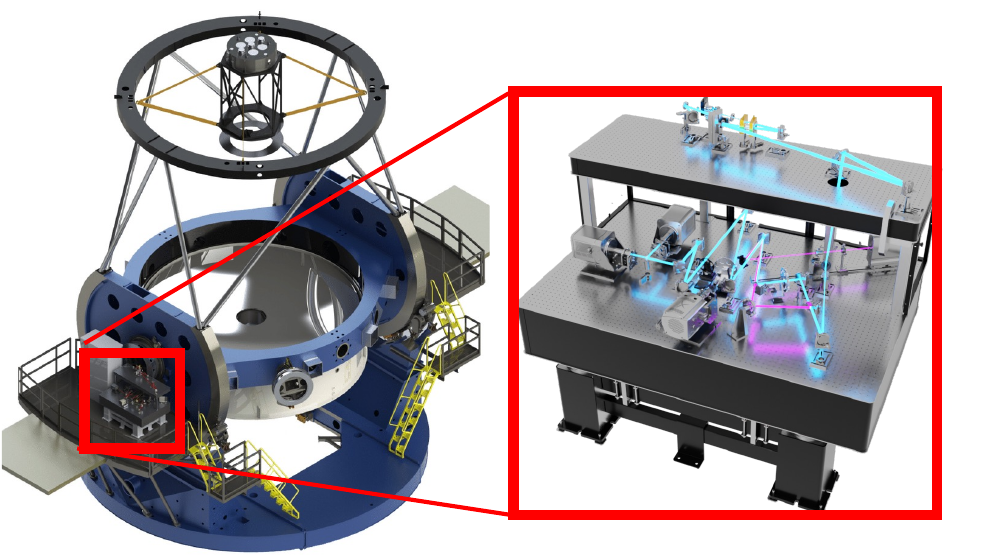}
   \end{center}
   \caption[example] 
%>>>> use \label inside caption to get Fig. number with \ref{}
   { \label{fig:telescope} 
A rendering of the 6.5 m Magellan-Clay telescope with MagAO-X installed on the Nasmyth platform highlighted and enlarged.}
   \end{figure}

MagAO-X \cite{males_magaox_2018,close_optical_2018} is the ``extreme" coronagraphic adaptive optics (AO) instrument for the 6.5-meter Magellan Clay telescope (Figure \ref{fig:telescope}) which is optimized to image at visible wavelengths. The instrument includes an array of deformable mirrors (DM) including: a high-stroke 97 actuator woofer ALPAO \textit{DM97}, a 2040 actuator tweeter DM (Boston Micromachines; BMC), and an additional DM for non-common path correction (NCPC) in the coronagraph.

The woofer-tweeter DM architecture\cite{brennan_advanced_2006} is driven by a $\sim 3$ kHz transmissive pyramid wavefront sensor. The $50\times50$ 2040-actuator tweeter provides high order wavefront correction with low-order modes offloaded to a $11\times11$ 97-actuator woofer DM.

One unique aspect of MagAO-X is its additional DM for NCPC and Low-Order Wavefront Sensing (LOWFS) within the coronagraph arm which we recently swapped with a 1000 actuator BMC \textit{Kilo-DM} as part of the MagAO-X "Phase II" upgrades\cite{males_magaox_2022}. This new NCPC DM surface required characterization to confirm device quality and for commanding the reflective face sheet to a flat state for efficient installation.

The main scientific motivation for the abovementioned Phase II upgrades is to significantly enhance MagAO-X's ability to image reflected starlight at close inner-working angles (IWAs). Currently, one of the main limitations for this mode of direct imaging is noise due to non-common path aberrations (NCPAs) which are instrumental in origin and originate downstream of the wavefront sensor (WFS) arm, so the AO system is unable to correct for them in typical configurations. Furthermore, the 
``quasi-static" speckles due to NCPAs are difficult to remove through conventional post-processing methods (e.g., angular differential imaging\cite{marois_angular_2006}) because they evolve too slowly to be averaged and too quickly to be removed using a single speckle pattern realization for a typical science observation. After characterization and installation of the Kilo-DM, we used it to address these concerns. Specifically, this upgrade led to faster LOWFS correction speeds and facilitated the Focus-Diversity Phase Retrieval (FDPR) and implicit Electric Field Conjugation (iEFC) algorithms on-sky for quasi-static speckle removal and in-situ Strehl ratio optimization, respectively.

\section{DM Characterization}

We replaced MagAO-X's NCPC DM as part of the Phase II upgrades. Namely, we swapped the old ALPAO DM97 with a new BMC Kilo-DM which features a $32 \times 32$ 952 microelectromechanical system (MEMS) actuator grid rated for $3.5 \mu$m of peak-to-valley (PV) surface stroke. Before the DMs could be exchanged, the Kilo-DM needed to be characterized to avoid aberrating MagAO-X's point-spread function (PSF) and maintaining accurate wavefront control.

\subsection{Procedures}

   \begin{figure}
   \begin{center}
   \includegraphics[width=0.33\linewidth]{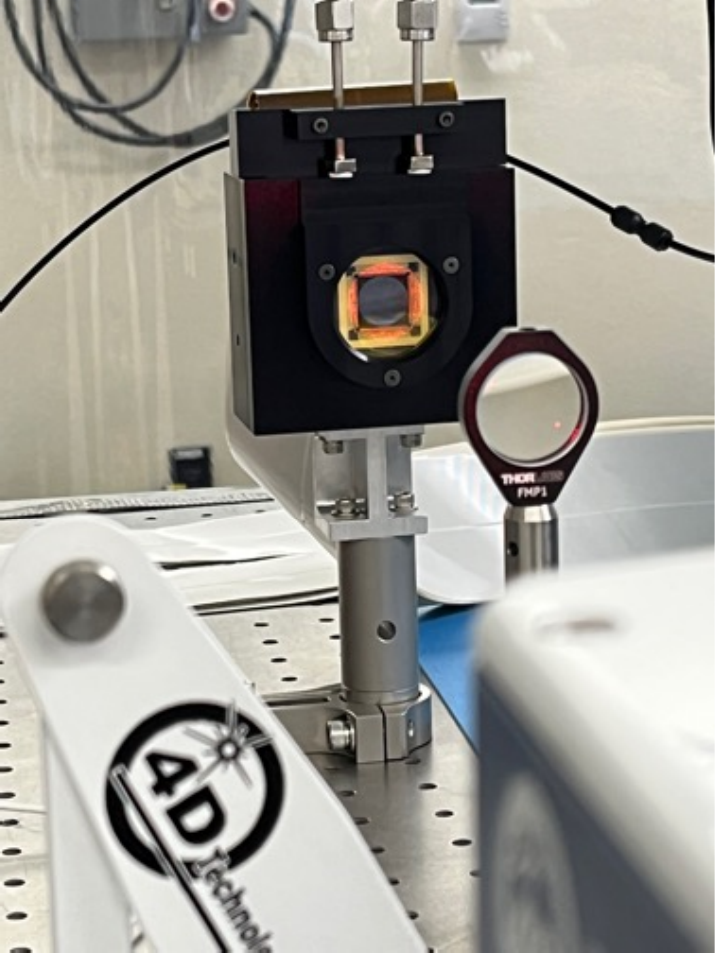}
   \includegraphics[width=0.33\linewidth]{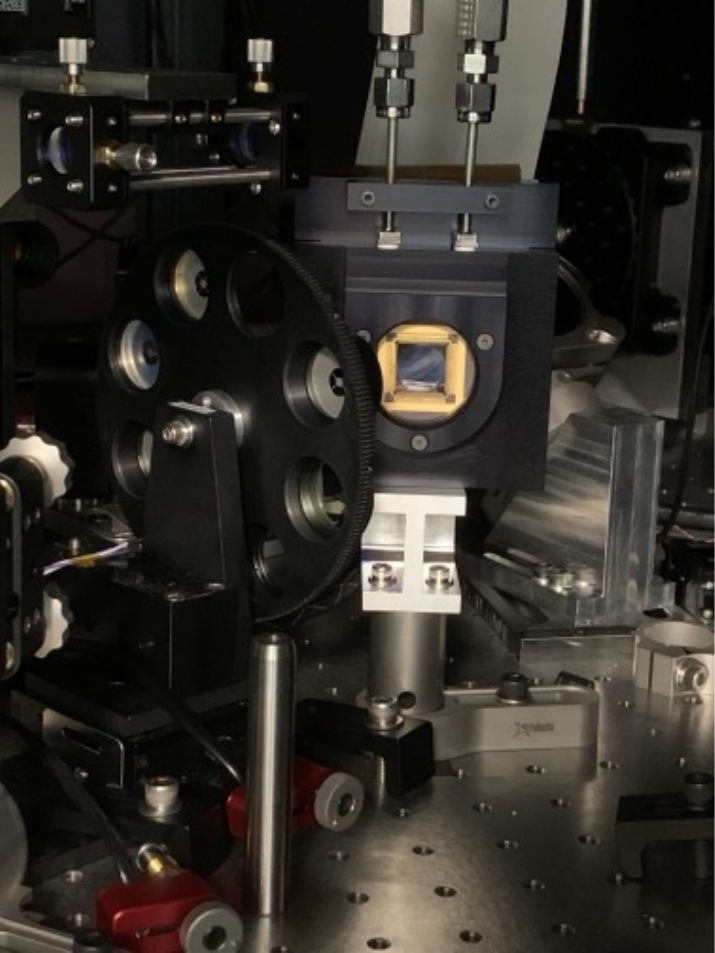}
   \end{center}
   \caption[example] 
%>>>> use \label inside caption to get Fig. number with \ref{}
   { \label{fig:dm_pics} 
The BMC Kilo-DM (left) set up on the testbed and (right) installed in MagAO-X after characterization.}
   \end{figure}

\begin{figure}
   \begin{center}
   \includegraphics[width=0.85\linewidth]{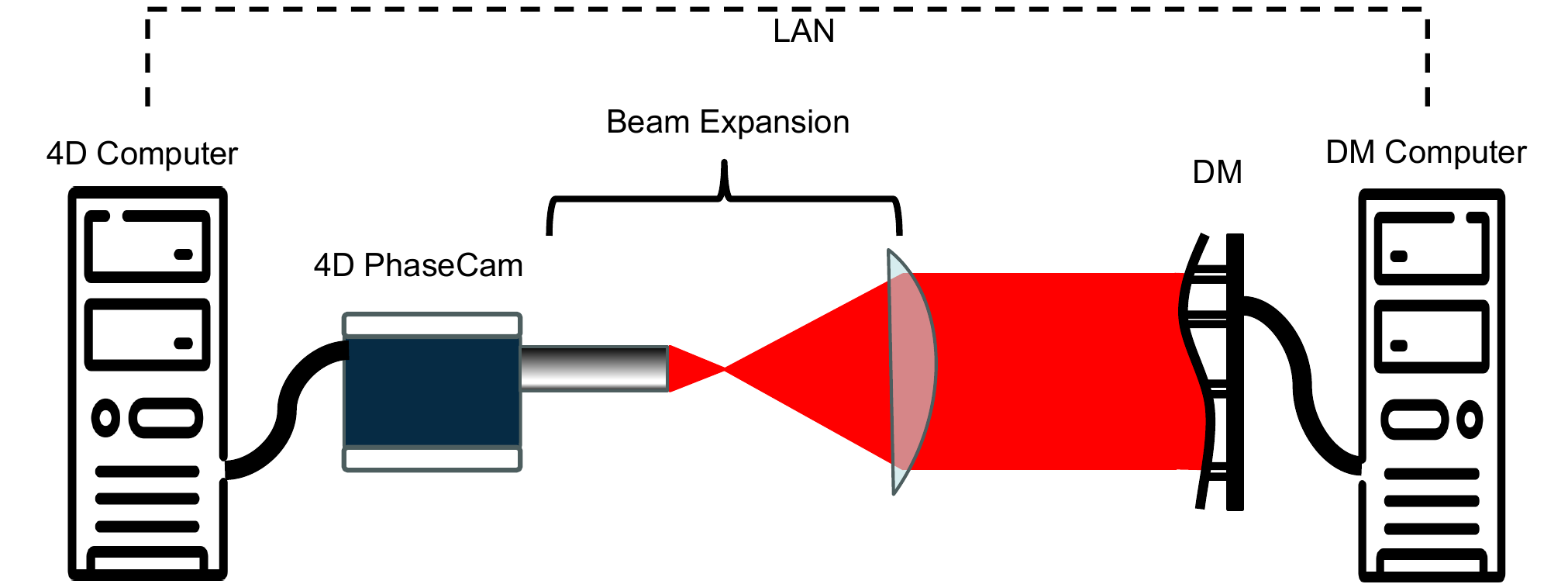}
   \end{center}
   \caption[example] 
   { \label{fig:opti_layout} 
Optical setup we utilized for measuring DM influence functions for driving the surface flat in closed-loop.}
   \end{figure}

We set up and aligned the DM on a test bench such that the surface could be measured in real time with a 4D Technology \textit{PhaseCam} (Twyman-Green) interferometer (Figure \ref{fig:dm_pics}) utilizing a HeNe laser source ($\lambda = 632.8$ nm). We used the BMC-supplied API to command the DM and the Python API provided by 4D Technology for interferometer operation, post-processing, and writing the measurements to disk. Because the diameter of the stock source beam was not large enough to measure the full test surface in a single exposure, we made use of additional lenses to expand the beam size. These expanding optics introduced low-order aberrations which were subtracted in post-processing through the use of a reference made from a 4” flat mirror ($\lambda/20$ PV; Figure \ref{fig:opti_layout}).

We built an interaction matrix to command the DM surface using the same general procedures established during past characterization of MagAO-X's tweeter and woofer DMs\cite{vangorkom_characterizing_2021}. Namely, we first commanded a global DM surface command to a voltage bias of 70\% to maximize single actuator stroke then measured single-actuator IFs by recording positive and negative DM pokes with amplitude $0.16$ $\mu$m ($\lambda/4$). We then computed the difference image to remove the surface form at its bias position; see Van Gorkom et al. \cite{vangorkom_characterizing_2021} for a representative IF for a MEMS DM. To expedite measurements of these IFs, we built a Python-based pipeline to automate DM and interferometer operations supported by a simple file monitoring class. On both the DM and 4D computer, the monitor process passively watched for changes to an empty file before allowing a DM command or interferometer capture to take place. For either a successful DM command or interferometer measurement, the pipeline then updated the empty file on the conjugate computer and the process continued, in sync, until all measurements were written to disk.

After capturing the IFs, we flattened and stacked the images to make the response matrix. Put mathematically,

\begin{equation}
    \vec{s} = F\vec{a}
\end{equation}

where $\vec{s}$ is a vector portraying the form of the DM surface, $F$ is the response matrix, and $\vec{a}$ is a vector of actuator commands. Furthermore, the backbone for the closed-loop flattening operation is projection of the measured DM surface onto the interaction matrix, which is given by the pseudo-inverse of $F$, to obtain an approximation in DM space:

\begin{equation}
    \vec{a} = F^{\dagger}\vec{s}
\end{equation}

where $F^{\dagger}$ denotes the pseudo-inverted response matrix. To summarize, we assume that we can approximate an arbitrary surface shape $\vec{s}$ given a sufficient span of orthogonal IF measurements $F$. Since we measured the IFs of single-actuator pokes, no IF can be reconstructed from any combination of IFs from the other actuators thus this condition is satisfied.

\begin{figure}[htbp]
   \begin{center}
   \includegraphics[width=0.35\linewidth]{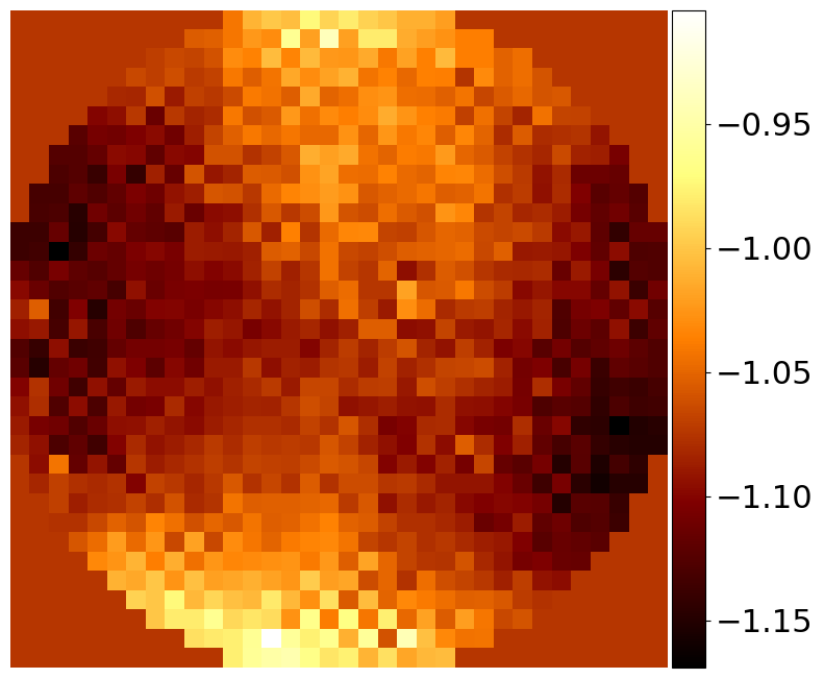}
   \includegraphics[width=0.35\linewidth]{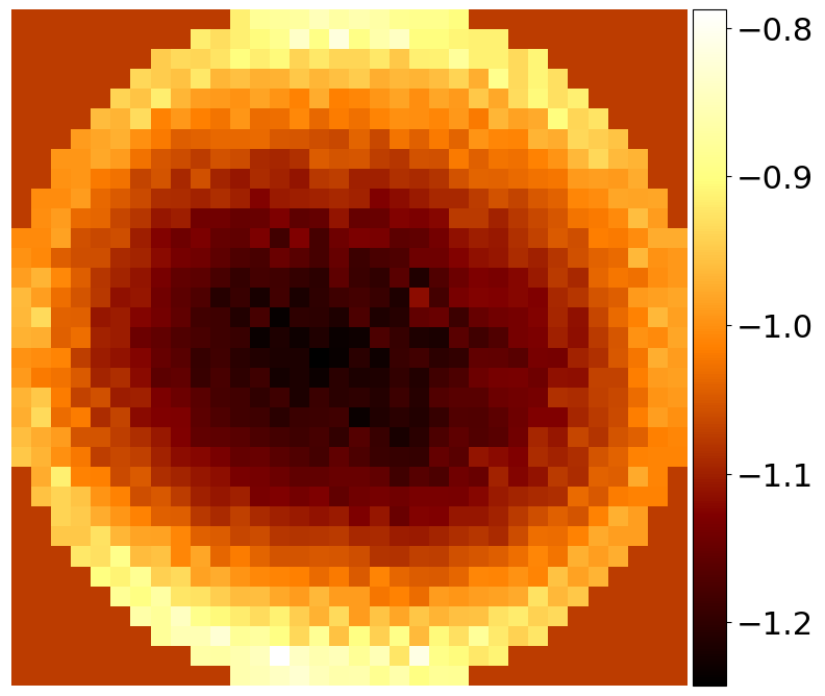}
   \end{center}
   \caption[example] 
%>>>> use \label inside caption to get Fig. number with \ref{}
   { \label{fig:flats} The best-fit DM command maps after the closed-loop flattening procedure. The colorbars are in units of microns. Left: The command map after 12 iterations pertaining to the "relaxed" flat where tip/tilt and defocus was removed prior to calculating the surface rms about the mean. Right: The command map after 11 iterations for the "absolute" flat where only tip/tilt was removed in post-processing.}
   \end{figure}

Minimizing the surface rms about the mean, we used the abovementioned interaction matrix $F^{\dagger}$ to drive the surface to a flat state, in closed-loop, using the  interferometer measurements (i.e., the WFS) to obtain $\vec{s}$. We found solutions to two types of flats: a ``relaxed" flat where tip/tilt and defocus aberrations were fit and removed in post-processing and thus were ignored when evaluating the fit metric and an ``absolute" flat where only tip/tilt was ignored (Figure \ref{fig:flats}).

\begin{figure}[ht]
   \begin{center}
   \includegraphics[width=0.25\linewidth]{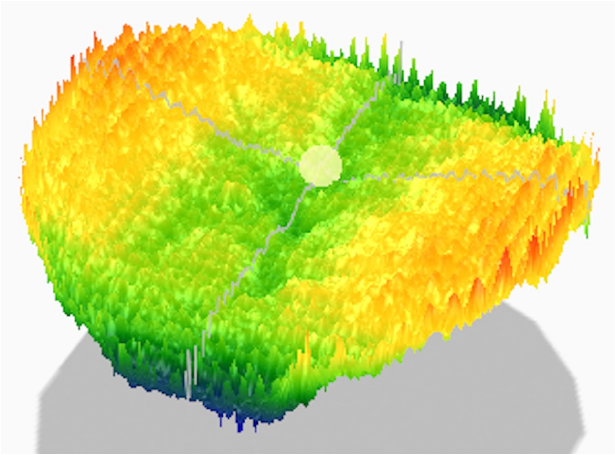}
   \includegraphics[width=0.25\linewidth]{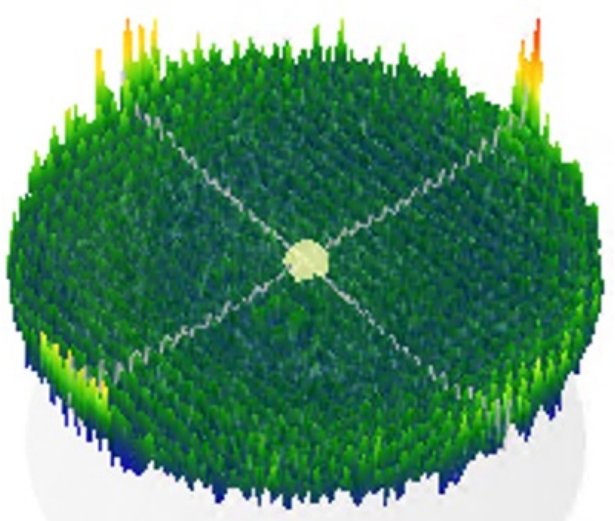}
   \end{center}
   \caption[example] 
%>>>> use \label inside caption to get Fig. number with \ref{}
   { \label{fig:surfs} 
DM surface pre- and post-flattening. Left: The bias position with defocus removed exhibited roughly 47 nanometers surface rms. Right: The surface rms reduced to about 10 nanometers, post-flattening. The spikes along the edges of the surface map correspond to edge effects leaking through the imperfect software mask and are $\sim 50$ nm above the surface mean.}
   \end{figure}

The relaxed flat converged in 12 iterations and the absolute flat converged in 11. We chose the relaxed flat to maximize the available stroke across all actuators\cite{vangorkom_characterizing_2021} and opted to instead compensate for the remaining defocus by translating MagAO-X's science and LOWFS cameras as well as the focal plane mask along the $z$-axis. The DM surface rms at the bias position was $\sim47$ nm and was reduced to $\sim10$ nm after the flattening procedure (Figure \ref{fig:surfs}).

\subsection{Installation and Optimization}

 \begin{figure}[htbp]
   \begin{center}
   \includegraphics[width=0.6\linewidth]{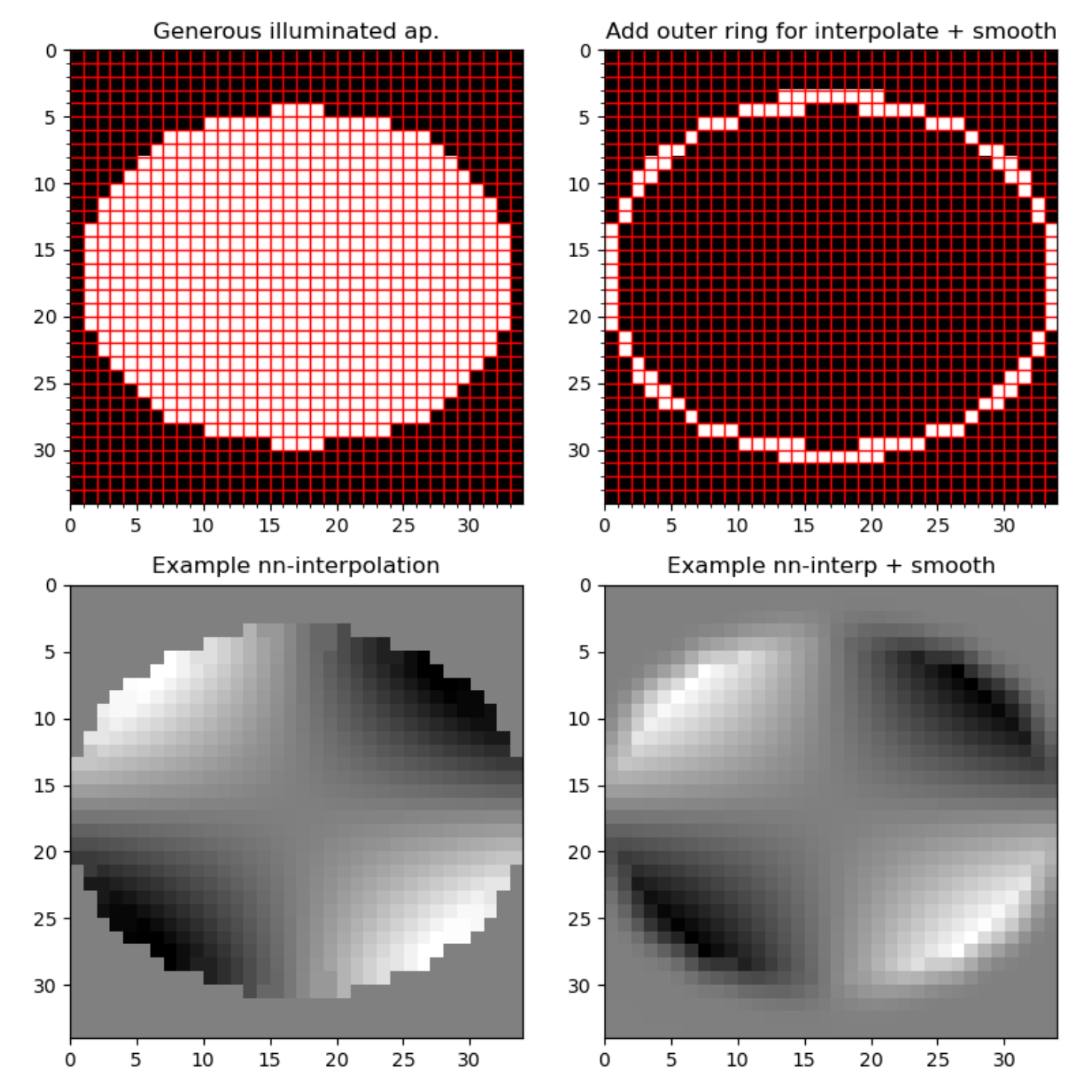}
   \end{center}
   \caption[example] 
   { \label{fig:basis} 
To assemble a basis for use on MagAO-X’s new NCPC DM, we (1, top left) define a generous aperture of size 32 by 26 actuators, (2, top right) isolate an outer single actuator-wide ring to ensure that there are no sharp surface features at the edge of the aperture, (3, bottom left) perform nearest-neighbor interpolation within this outer ring, and (4, bottom right) apply Gaussian smoothing. All values exterior to the original aperture from (1) are replaced with the interpolated + smoothed values. An astigmatism component is shown for example purposes.}
   \end{figure}
   
After obtaining the best-fit command map for a relaxed flat, we installed the new NCPC DM in MagAO-X in late 2023 (see the right panel in Figure \ref{fig:dm_pics} to see the new DM installed in the instrument). We then aligned it in the lab using the internal telescope simulator and a commanded binary pattern marking the shape and position of the expected beam footprint on the NCPC DM. However, we observed some remaining static aberrations even with the best-fit flat command. To scrub out these residual aberrations, and for general utility, we constructed an orthogonal basis for use over the area of beam footprint using Zernike modes up to Z51. In practice, this basis is useful for manually dialing out or inducing specific aberrations at the detector focal plane or for use with, e.g., the ``eye-doctor"\cite{bailey_lbt_2014,vangorkom_characterizing_2021} algorithm for automated instrument Strehl ratio optimization. Figure \ref{fig:basis} details the process we used to construct such a basis. We note that due to the highly elliptical illuminated area on the DM, the question of how best to control the actuators outside of this region is still being actively investigated.

\section{On-Sky Performance}

MagAO-X went on-sky with the new NCPC DM during the observing block in March---May 2024, spearheading enhancements to MagAO-X’s unique ability to do coronagraph-stage, real-time wavefront control. We had ample opportunity to test the upgraded configuration during engineering time and science observations; the results of these tests are briefly discussed in the next few subsections.

   \begin{figure}
   \begin{center}
   \includegraphics[width=0.6\linewidth]{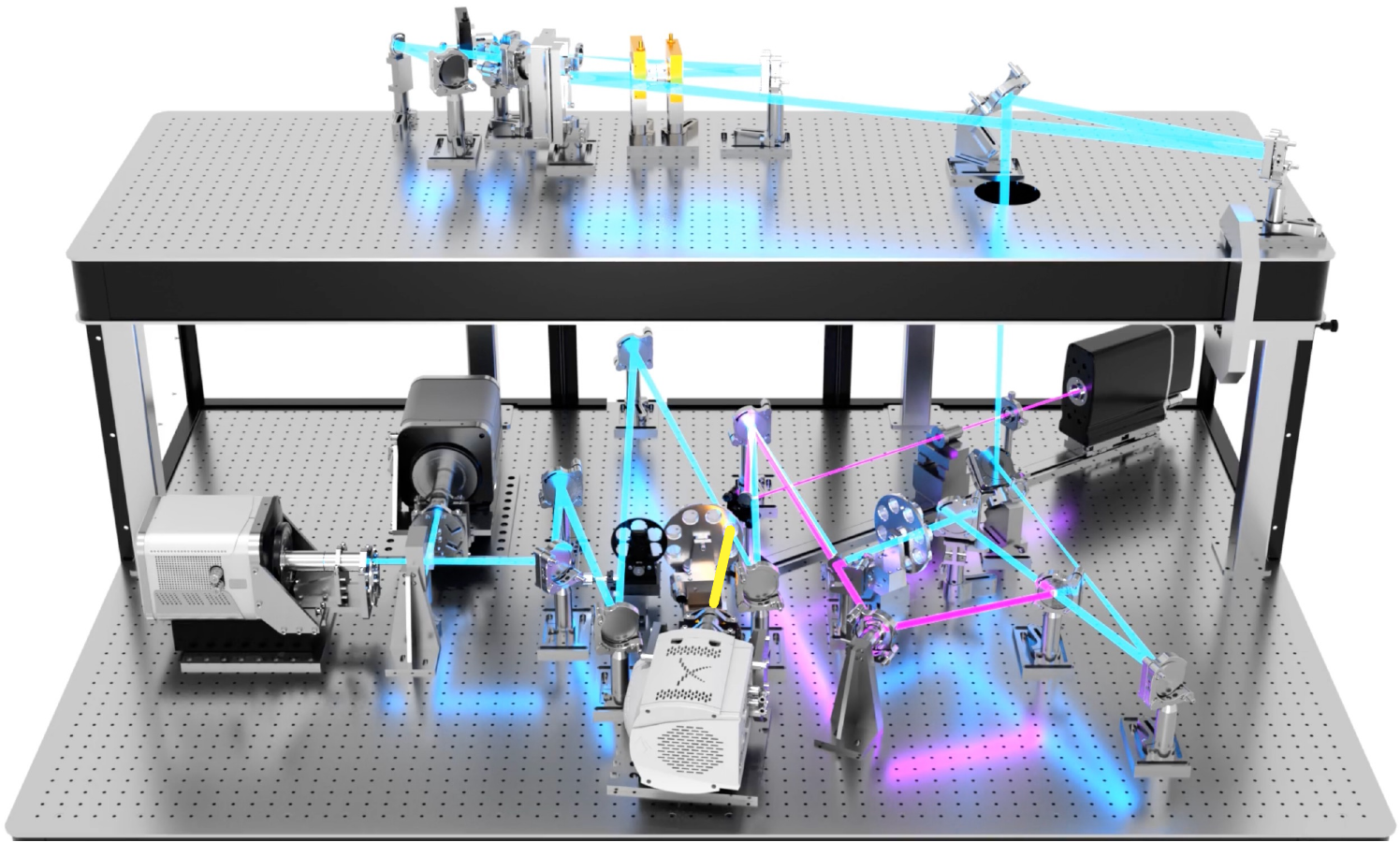}
   \end{center}
   \caption[example] 
   { \label{fig:beams} 
A render of MagAO-X tracing the beam paths as they traverse the upper and lower optical benches. The blue beam starts on the upper bench and shows the path taken by the light from the telescope. The magenta beam illustrates the beam to the wavefront sensor. The yellow beam shows the light reflected off the coronagraph for LOWFS.}
   \end{figure}

\subsection{NCPC and LOWFS}

NCPAs originate within the instrument due the disconnect between the WFS and science arms. Since NCPAs are not sensed by the high-order WFS, they become a limiting factor for the contrast at the imaging plane. One way to fight this phenomenon is adding additional WFSs further along in the optical train. 

MagAO-X makes use of light rejected by the reflective coating of the Lyot focal plane mask for LOWFS (yellow beam in Figure \ref{fig:beams}) for NCPA correction \cite{mcleod_thesis_2023}. In addition to a considerable increase in the number of modes LOWFS can correct for, the new Kilo-DM also increased the maximum closed-loop correction speeds (from 2kHz to 10 kHz), increasing robustness to instrumental jitter. We plan to perform more comprehensive tests with 10kHz correction speeds as well as measuring and calibrating mode linearity with LOWFS during the next MagAO-X observing block in Fall 2024. 

\subsection{Mitigating quasi-static speckles with iEFC}

\begin{figure}
   \begin{center}
   \includegraphics[width=0.6\linewidth]{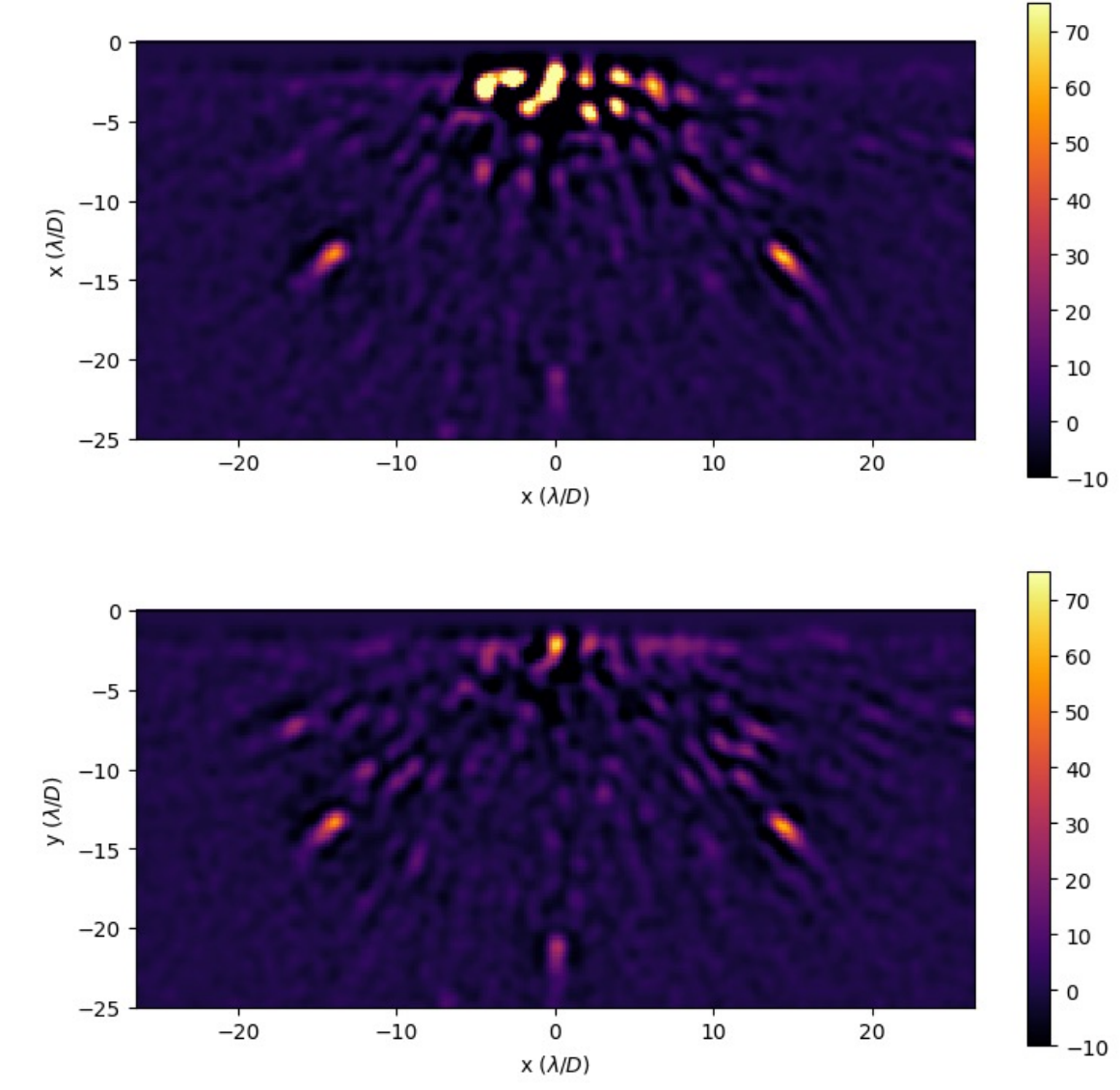}
   \end{center}
   \caption[example] 
   { \label{fig:iEFC} 
iEFC results on the star $\beta$ Corvi (V = 2.6 mag). The star is located at the top of both images and is blocked with a knife-edge coronagraph. Top: iEFC disabled and quasistatic speckles are readily seen. Bottom: Marked decrease in quasistatic speckles with iEFC enabled. The bright dot remaining at the IWA is an injected companion. The prominent speckles that are present in both images at the $\sim4$, 6, and 8 o’clock positions are induced by the DM surface.}
   \end{figure} 

Quasistatic speckles are a source of noise that stems from NCPAs and are debilitating for AO imaging performance at close IWAs\cite{hinkley_temporal_2007}. During the last MagAO-X observing run in 2024A, we demonstrated focal plane wavefront sensing, on-sky, using the iEFC algorithm\cite{haffert_advanced_2022} to mitigate this noise source. Briefly, this algorithm combines Electric Field Conjugation\cite{giveon_broadband_2007} with an empirical calibration step via the NCPC DM (i.e., no optical model is needed to reconstruct the electric field). With knowledge of the morphology of the electric field, the NCPC DM can then cancel by instilling the opposite electric field. Our on-sky iEFC efforts produced a rectangular dark hole of size $11 \times 13$ $\lambda/D$ which is a region of high-contrast almost completely devoid of quasistatic speckles (Figure \ref{fig:iEFC}). These results will be described in detail in a later publication (Haffert et al., in prep). One avenue for future work is studying the effects of DM registration to the detector plane.

\subsection{In-situ Strehl ratio optimization with FDPR}
   
   \begin{figure}
   \begin{center}
   \includegraphics[width=\linewidth]{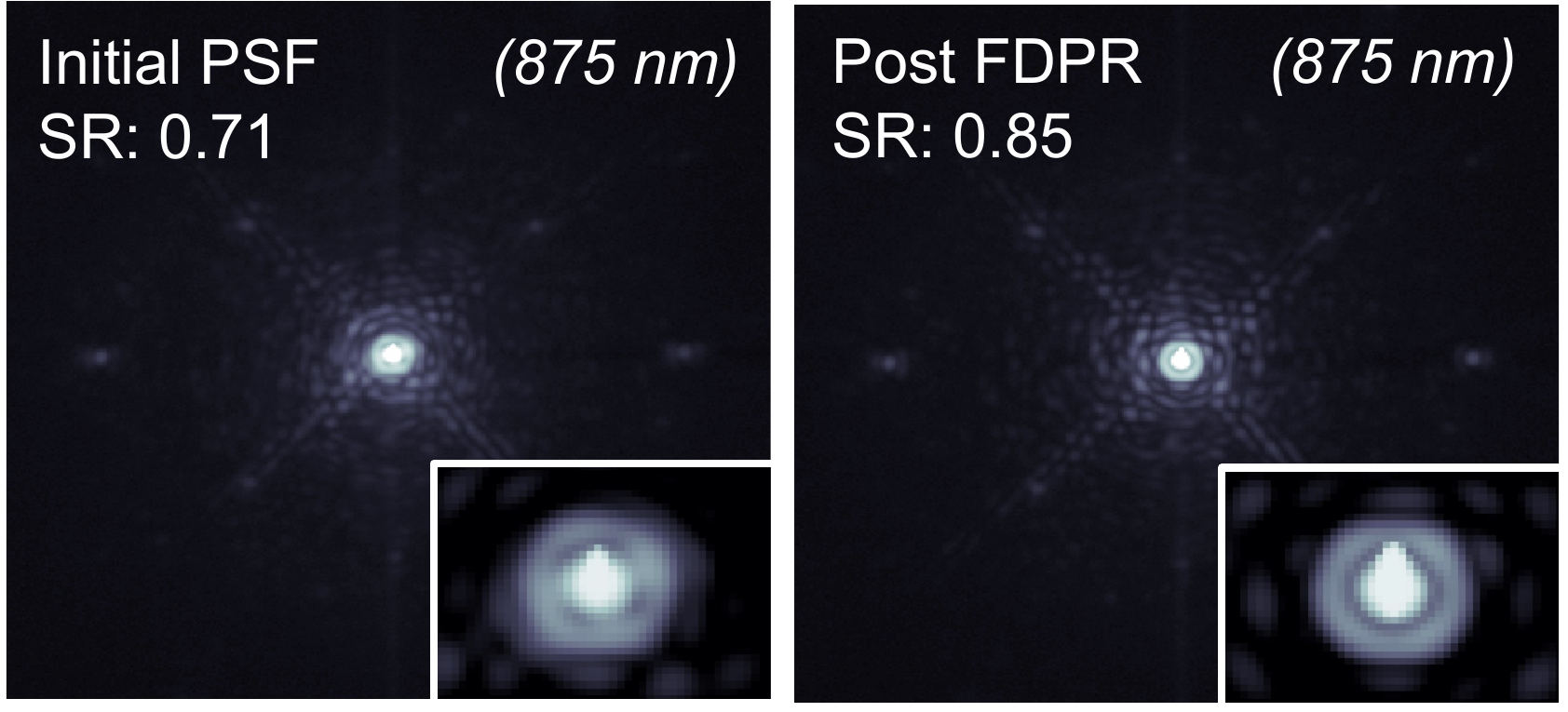}
   \end{center}
   \caption[example] 
   { \label{fig:fdpr} 
Closed-loop FDPR results on the star $\beta$ Corvi (V = 2.6 mag) imaged through a narrowband $CH4$ filter (875 nm). The initial Strehl ratio was 0.71. After 10 iterations of closed-loop FDPR, the Strehl ratio increased to 0.85. The PSF core is shown inset in the bottom right corner.}
   \end{figure}

The FDPR algorithm\cite{thurman_amplitude_2009,fienup_phase_2013,vangorkom_characterizing_2021} is one way to measure the spatial distributions of phase and amplitude at the pupil plane in an AO system. At a high level, FDPR attempts to minimize the error between PSF models and images at multiple planes of defocus to avoid solution degeneracies.

We report successful on-sky use of closed-loop FDPR leading to drastic improvements in Strehl ratio as part of an initial calibration step before observing a given science target. To supply the algorithm with the necessary components, we use the NCPC DM to apply precise amounts of defocus to the PSF image which is recorded on one of MagAO-X's science cameras. After the requested amount of iterations is completed on a bright ($V \lesssim 5$ mag) star, we use the NCPC DM command map that minimizes differences between the PSF images and models as the new flat command for any subsequent science observations in a nearby portion of the sky. Figure \ref{fig:fdpr} shows the degree of improvement for the Strehl ratio after 10 iterations of closed-loop FDPR while imaging through a narrow $CH4$ filter (875 nm) on the bright star $\beta$ Corvis. We note, however, that FDPR does instantaneous band-limited measurements for the Strehl ratio thus neglecting vibrations; the true Strehl ratio is likely lower by about 0.1 to 0.2.

Currently, for on-sky use, the defocused PSF images are first averaged for $\sim 5$ seconds to temper the effects of atmospheric turbulence before feeding it into the algorithm. There would be value in further study as to determine optimal FDPR parameters based on the current observing conditions (e.g., how do we best average the PSF images over turbulence?).

\section{Summary and Future Work}

We upgraded MagAO-X's NCPC DM to a BMC Kilo-DM as part of a comprehensive suite of ongoing upgrades\cite{males_magaox_2022} to be completed in the next couple of years. We detailed successful characterization of the device prior to installation within the instrument. We also shared our on-sky results using the new NCPC DM during the MagAO-X observing block in 2024A. The key takeaways from our on-sky tests with this new DM are listed as follows:

\begin{itemize}
    \item MagAO-X's unique ability to do coronagraph-stage wavefront control has been greatly improved.
    \item We demonstrated quasi-static speckle removal in a $11 \times 13$ $\lambda/D$ control radius "dark hole" without any loss of stroke from the 2040-actuator tweeter DM.
    \item The FDPR algorithm is an effective way to drastically improve Strehl ratio as part of an initial calibration step before observing a given science target.
    \item LOWFS on MagAO-X is much improved with faster correction speeds, increasing robustness to instrumental jitter.
\end{itemize}

\acknowledgments 
We are thankful for support by the NSF MRI Award no. 1625441 (MagAO-X). This work and the rest of the MagAO-X Phase II upgrades are made possible by the generous support by the Heising-Simons Foundation.

\bibliography{report} 
\bibliographystyle{spiebib} 

\end{document}